\documentclass[aip,graphicx]{revtex4-1}

\draft 
\usepackage{graphicx}
\usepackage{textcomp}
\usepackage{subcaption}
\usepackage{braket}
\usepackage{bbold}
\usepackage{amsmath}

\begin{document}


\title{Entanglement and coherence in pure and doped Posner molecules} 

\author{Betony Adams}
\affiliation{Quantum Research Group, University of KwaZulu-Natal}
\affiliation{National Institute for Theoretical and Computational Sciences, South Africa}
\affiliation{The Guy Foundation, Dorset, UK}
\author{Ilya Sinayskiy}
\affiliation{Quantum Research Group, University of KwaZulu-Natal}
\affiliation{National Institute for Theoretical and Computational Sciences, South Africa}
\author{Shivang Agarwal}
\affiliation{The Department of Electrical and Computer Engineering, University of California Los Angeles, CA, 90095, USA}
\author{Francesco Petruccione}%
\affiliation{Quantum Research Group, University of KwaZulu-Natal}
\affiliation{National Institute for Theoretical and Computational Sciences, South Africa}
\affiliation{School of Data Science and Computational Thinking and Department of Physics, Stellenbosch University, South Africa}
\date{\today}

\begin{abstract}
The potential role of spin in biological systems is a primary topic in quantum biology. However, much of this research focuses on electron spin. A recent hypothesis suggests that nuclear spin may be better suited to biological processes, being less sensitive to decoherence.  The hypothesis details how phosphorus nuclei might be prepared in a spin entangled state, how this entanglement is protected by assembly into calcium phosphate (Posner) molecules, and how this entanglement might modulate calcium ion production and concomitant neural activation. In this paper we investigate the robustness of quantum effects such as coherence and entanglement in Posner molecules. We investigate how these effects are directly dependent on specific parameters such as spin-spin coupling strengths and Posner molecule symmetry. We also investigate how lithium isotope doped Posner molecules differentially modulate quantum resources such as coherence and entanglement and whether this is a viable explanation for lithium's mechanism of action in bipolar disease. Finally we illustrate how entanglement might possibly be preserved through exploitation of the biological environment.
    \end{abstract}
\maketitle 
\section{\label{sec:level1}Introduction}
The role of quantum spin in biological systems is one of the primary topics of quantum biology \cite{kim,marais2018,alkhalili}. A pivotal question in the field is whether quantum coherence -- and possibly entanglement -- may contribute to the functional importance of biological processes, such as the charge transfer that underpins photosynthesis or the spin-dependent chemical reactions that may constitute the mechanism of the avian compass. Spin chemistry has a long history, with its origins in the 1960s \cite{steiner,horespin}. Spin biology, in which spin-dependent chemical reactions are modulated by the geomagnetic field, has almost as long a history. In the 1970s it was first proposed that the radical pair mechanism might explain how birds navigate so accurately across great distances \cite{schulten78}. The radical pair mechanism focuses on the spin states of paired electrons. More recently, however, it has been suggested that nuclear spin is better suited to playing a role in biological processes, due to the fact that nuclear spin has much longer coherence times than electron spin \cite{fisher1,fisher2,fisher3}. In particular, it has been suggested that phosphorus nuclear spin, bound into calcium phosphate molecules known as Posner molecules, is ideally suited to play a role in cognition and memory. Spin half phosphorus nuclei have very long decoherence times, a factor that is increased by their `shielding' in Posner molecules by spin-zero calcium nuclei \cite{fisher1,fisher2,fisher3}.\\ 
\\
Quantum to biological transduction is an integral step in the modelling of quantum effects in biological processes. In the radical pair mechanism this step is achieved by the electronic spin selectivity of the chemical reactions involved in biological functioning \cite{fay,hore2016,rodgers}. In the Posner molecule model, it has been suggested that the binding and hydrolysis of Posner molecules is dependent on the phosphorus nuclear entanglement \cite{fisher1}. Hydrolysis of Posner molecules releases free calcium ions, which in turn play a powerful role in cellular and neural signalling. In this way quantum entanglement is implicated in coordinated calcium signalling \cite{fisher1}. In this paper we take a closer look at the quantum properties of Posner molecules, such as coherence and entanglement. Given that there are six phosphorus nuclei in each Posner molecule, we investigate the spin interactions and how these influence coherence and entanglement. To begin with we use established measures of coherence and concurrence to quantify these quantum resources in pure Posner molecules, by which we mean those molecules that only contain calcium and phosphate ions. We then extend this model to look at the common phenomenon of doped Posner molecules, in which spin zero calcium ions are replaced by other ions with nuclear spin, such as lithium and hydrogen. In particular we look at the effects of lithium ion substitution on coherence and entanglement. Our motivation for investigating lithium is the evidence that different lithium isotopes have different behavioural effects in animal studies \cite{sechzer,ettenberg}. Lithium is an important drug in the treatment of bipolar disease. It has been suggested that the different outcomes of lithium treatment could be explained by the different nuclear spins of lithium isotopes, which, when incorporated into Posner molecules, would differently modulate the spin dynamics \cite{fisher1,fisher2}. We model the spin dynamics in order to confirm whether different lithium isotopes have different effects on entanglement and thus Posner hydrolysis and free calcium ions. In order to determine whether it is only the  spin entanglement that is instrumental in the different effects of lithium ions we investigate other ways in which lithium isotopes may differ in their mode of action. In particular we look at the role of spin in relaxation mechanisms and whether the different isotopes induce spin relaxation to different degrees. We also apply results from radical pair literature, such as the effects of electromagnetic noise of different frequencies, to gain insight into nuclear spin dynamics. And finally, we investigate a novel way in which entanglement between phosphorus nuclei, before they are bound into Posner molecules, is enhanced rather than destroyed by interaction with hydrogen ions.

\subsection{The radical pair mechanism: electron spin}
 The radical pair model of avian magnetoreception has given rise to a number of papers outlining how birds might sense the Earth's magnetic field through the spin states of paired electrons. This radical pair compass can be outlined in the following three steps. First an incident photon transfers its energy to a donor molecule causing electron transfer, which results in a spatially separated electron pair that is conventionally taken to be in a singlet state. Second, under the influence of the geomagnetic Zeeman effect and the hyperfine interaction with surrounding nuclei, the spin state begins to interconvert between singlet and triplet states. Finally, recombination occurs, which is dependent on the spin state \cite{fay,hore2016,rodgers}. In this way spin states translate into biologically relevant signalling states. \\
\\
The role that spin coherence might play in the sensitivity of this magnetic compass has been investigated in a number of papers \cite{ganguly,kattnig,le}. There has also been discussion of the importance of quantum entanglement in this mechanism \cite{cai,gauger,pauls,zhang}. This follows from the fact that the radical pair is conventionally taken to originate in a maximally entangled singlet state. For two spin-half electrons, if the first electron of the pair is in a state given by the vector
\[
\ket{\uparrow}=
	\begin{pmatrix} 1 \\ 0 \end{pmatrix},
\]
and the second electron is in the state given by the vector
\[
\ket{\downarrow}=
    \begin{pmatrix} 
    0 \\
    1 
    \end{pmatrix},
\]
then the singlet state can be written as:
\begin{equation}
S=\frac{1}{\sqrt{2}}(\ket{\uparrow\downarrow}-\ket{\downarrow\uparrow}).
\end{equation}
However, once this singlet is exposed to the nuclear environment it can undergo conversion to the triplet states, only one of which is entangled, 
\begin{equation}
T_0=\frac{1}{\sqrt{2}}(\ket{\uparrow\downarrow}+\ket{\downarrow\uparrow}).
\end{equation}
While the subject is still up for debate, there is some indication that, while coherence is important, entanglement is not strictly necessary for the radical pair mechanism to function \cite{gauger,hogben,kattnig}. What is potentially interesting, however, is that should entanglement play a role in this mechanism, one way in which to enhance the entanglement lifetime is the application of appropriate magnetic fields. This would create a maximally entangled subspace by increasing the energy separation of the two non-entangled triplet states $T_+=\ket{\uparrow\uparrow}$ and $T_-=\ket{\downarrow\downarrow}$ \cite{tiersch}. In this paper we investigate coherence and entanglement in a model that is analogous to the radical pair mechanism but utilises correlated nuclear spins rather than correlated electron spins.
\subsection{The Posner molecule mechanism: nuclear spin}
Calcium ions are integral to physiological processes in the body. They play an important role in, among other things, the release of neurotransmitters and the activation of nerve cells \cite{sudhof,neher}. Amorphous calcium phosphate has been proposed as an essential reservoir for calcium ions in biological systems \cite{wolf}. One of the forms that calcium phosphate can take is the Posner cluster, first identified by Betts and Posner and referred to in this paper as the Posner molecule \cite{posner}. A recent hypothesis suggests that quantum effects could play a role in Posner molecule formation and dissolution and thus also the storage and release of calcium ions. In this way, entanglement of spin-half phosphorus nuclei that are subsequently assembled into Posner molecules may play a role in quantum cognition \cite{fisher1}. Phosphorus nuclei make for good qubits due to the fact that spin-half nuclei have no quadrupole moment and longer relaxation times \cite{fisher1}. In Fisher's initial hypothesis, entangled phosphorus nuclei are achieved through the enzyme-catalysed hydrolysis of pyrophosphate, which consists of two phosphates \cite{fisher1}. Pyrophosphate is a byproduct of adenosine triphosphate (ATP), a molecule integral to energy processes in cells. It has been suggested that due to spin constraints on the molecular dynamics of a pyrophosphate molecule bound to the enzyme pyrophosphatase, the two phosphates are produced with their phosphorus nuclei in a singlet (maximally entangled) state. If these entangled phosphates are bound with calcium into separate Posner molecules, then this might be thought of as creating entangled Posner molecules, where the phosphorus spin is shielded from decoherence by a spin-zero cage of oxygen and calcium \cite{fisher1}. Furthermore, this entanglement may influence Posner molecule dynamics and future binding, which causes the molecules to melt and release calcium, which then has an effect on neural activation. In this way phosphorus nuclear spin entanglement might play a role in neural excitability. It has also been suggested that Posner molecules can be doped with ions other than calcium, such as magnesium or lithium, and that the non-zero nuclear spin of these ions would change the spin dynamics of the phosphorus nuclei \cite{fisher1, fisher2,fisher3}. See Figure 1 for a schematic of undoped and doped Posner molecules. The relative composition of Posner molecules has been shown to depend on other ions available during their formation \cite{mancardi}. Two monovalent lithium ions, for example, might take the place of the central divalent calcium ion \cite{fisher1}. The non-zero spin of the lithium nuclei would then have a resultant effect on phosphorus spin dynamics and potential neural activation. This could also explain the interesting experimental result that lithium isotopes, $^{6}\textrm{Li}$ and $^{7}\textrm{Li}$, have differing effects on parenting behaviour as well as hyperactivity in rats \cite{sechzer,ettenberg}. With this in mind we model the spin dynamics of a pair of spin-correlated phosphorus nuclei in two Posner molecules dopes with different lithium isotopes. We use this model to investigate some of the questions raised by the Posner molecule model of neural entanglement: how might we measure nuclear coherence and entanglement, do different lithium isotopes have an effect on this entanglement, and what effects do these isotopes have on the spin dynamics?
\begin{figure*}[t!]
	\centering
	\includegraphics[scale=0.6]{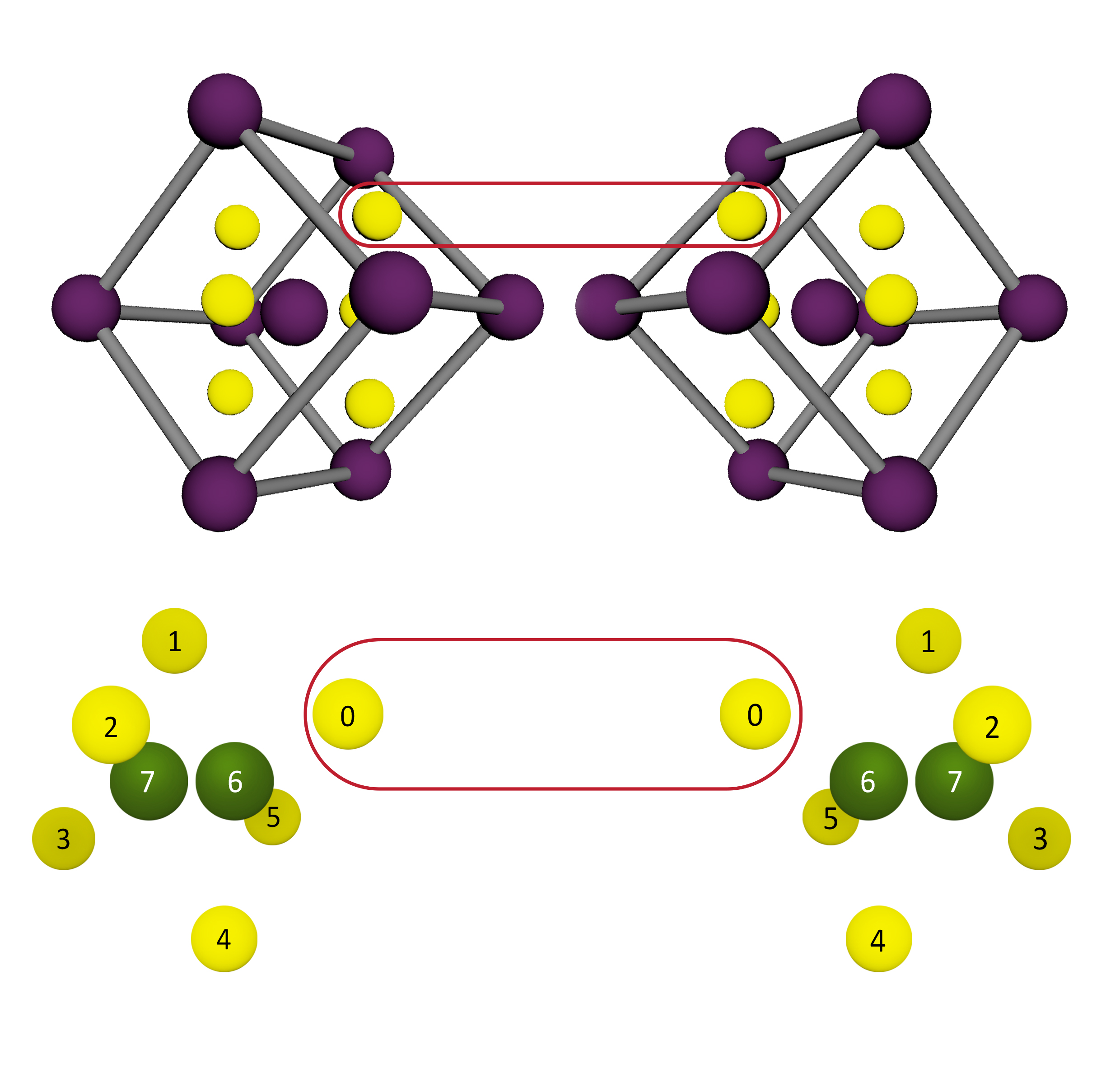}
	\caption{A schematic representation of undoped and doped Posner molecules. At the top are two Posner molecules, with eight calcium ions (purple) at the corners of the cube and one calcium ion at the centre. On each of the faces of the cube is a phosphate (yellow). The red loop represents one set of entangled phosphates, with correlated phosphorus nuclear spin. At the bottom, we have removed the calcium ions (no nuclear spin) to simplify our representation of the spin dynamics. We have also introduced two lithium ions (doped Posners) instead of the calcium ion at the centre. For clarity in our results section we have numbered the relevant spins 0--5 for the phosphorus spins and 6--7 for the lithium spins. This means that an interaction between spin 0 and 1 will have J-coupling strength $J_{01}$ and so forth. The initial entanglement in our model is between spins 0 and 0 in each Posner molecule, which are in a singlet state.}
\end{figure*}
\section{Theory}
\subsection{A simplified model}
In a recent paper Player and Hore investigate the claim made by Fisher \cite{fisher1} that the lifetime of the entangled phosphorus nuclear spins might extend to as long as 21 days \cite{player}. By considering intramolecular spin interactions rather than intermolecular spin interactions, as Fisher does \cite{fisher1}, they reach a much reduced, though still appreciably long, lifetime of 37 minutes. In their paper they use concurrence to illustrate the dependence of entanglement on the singlet character of the phosphorus nuclear spins \cite{player}. Following their example we construct a spin Hamiltonian for three different variations of a pair of entangled Posner molecules: no lithium, $^{6}\textrm{Li}$ (nuclear spin 1) and $^{7}\textrm{Li}$ (nuclear spin $\frac{3}{2}$). The coherent spin Hamiltonian is given by:
\begin{equation}\label{Ham1}
\hat{H_{S}} =  B_0\sum_{k}\gamma_k S^k_{z}  + \sum_{i<k} 2\pi J_{ik}\vec{S^i}\vec{S}^{k}.
\end{equation}
For pure Posner molecules: $k=1,...,N_P$. For Posner molecules doped with lithium $k=1,...,N_P,N_{P+1},...,N_P+N_L$. This gives $\gamma_k=\{ k \le N_P: \gamma_P \mid k>N_P: \gamma_L\}$,
where $\gamma_P$ and $\gamma_L$ are the gyromagnetic ratios of $^{31}\textrm{P}$ and the different lithium isotopes and $B_0$ is the geomagnetic field.
The first term represents the Zeeman interaction for the phosphorus and lithium nuclei. The second term represents the indirect spin coupling or scalar coupling between the various nuclei in the pure Posner molecule or lithium doped Posner molecule, where $J_{ik}$ is the strength of the coupling between nucleus $i$ and $k$. Here, 

\[
  S^k_\alpha = \left\{ \begin{array}{l}
    k \le N_P: S^P_\alpha \\
    k > N_P: S^L_\alpha
  \end{array}\right\},
\]
where  $S^P_\alpha$ and $S^L_\alpha$ are the spin operators for phosphorus and lithium, with $\alpha=x,y,z$.
\\
\\
It has been proposed that introduction of lithium into Posner molecules might exert its influence through the interaction of lithium nuclear spin with phosphorus nuclear spin \cite{fisher1}. To investigate how lithium isotopes change the quantum behaviour of Posner molecules we include measures for both coherence as well as entanglement. As a coherence measure we use the basis-independent coherence given by \cite{le}:
\begin{equation}\label{Coherence}
C_{BI}(\rho)=\text{log}_2d- S(\rho),
\end{equation}
where 
\begin{equation}
S(\rho)=-\mathrm{tr}[\rho\log_2\rho],    
\end{equation}
is the von Neumann entropy. $C_{BI}(\rho)$ is the relative entropy distance to the maximally mixed state with dimension $d$ and $\ket{i}$ representing each of the singlet and triplet states,
\begin{equation}
\mathbb{I}_d/d=\sum^{d-1}_{i=0}\frac{1}{d}\ket{i}\bra{i}.    
\end{equation} 
To measure entanglement we follow the example of Player and Hore \cite{player}, using concurrence as originally formulated by Wootters  \cite{wootters}, given by
\begin{equation}\label{Concurrence}
C(\rho)=\text{max}(0,\sqrt{\lambda_1}-\sqrt{\lambda_2}-\sqrt{\lambda_3}-\sqrt{\lambda_4}),
\end{equation}
where the $\lambda_i$ are the eigenvalues in decreasing order of the matrix 
\begin{equation}\label{Concurrence2}
\rho(\sigma_y\otimes\sigma_y)\rho^*(\sigma_y\otimes\sigma_y),
\end{equation} 
where $\rho^*$ is the complex conjugate of $\rho$ and $\sigma_y$ is the relevant Pauli matrix.\\
\\
Exactly how quantum effects such as coherence and entanglement mitigate bipolar disorder is unclear. For undoped Posner molecules, Fisher hypothesises that the rotational entanglement inherited from their spin entanglement modulates their chemical binding. Binding in turn allows Posner molecules to melt and release calcium which is implicated in neural activation \cite{fisher1}. A recent paper by Halpern and Crosson translates Fisher's ideas into quantum information formalism. Among other things the paper details how entanglement might increase molecular binding rates \cite{halpern}. Increased binding rates would mean increased calcium production, enhanced neurotransmitter release (Fisher specifies glutamate) and altered neural activity. Glutamate is an excitatory neurotransmitter \cite{zhou}. Bipolar disease may be related to calcium signalling \cite{warsh,berridge,harrison} as well as neural excitability. Neurons from patients with bipolar disorder are hyperexcitable and, furthermore, this hyperexcitability was reversed by lithium treatment \cite{stern,mertens}. With this in mind it is instructive to investigate what effect lithium might have on Posner molecule entanglement and thus neural excitability. And to what different degree $^{6}\textrm{Li}$ and $^{7}\textrm{Li}$ might attenuate this excitability.

\subsection{The (problem of) parameters}
Parameters for the Zeeman part of the coherent Hamiltonian are well defined. The gyromagnetic ratios of $^{31}\textrm{P}$, $^{7}\textrm{Li}$  and $^{6}\textrm{Li}$ in MHz.T$^{-1}$ are 17.24, 16.55 and 6.27, respectively \cite{bernstein,spencer}. We take the Earth's magnetic field to be $50\mu$T. The primary problem we encountered in modelling the spin system is the lack of definitive J-coupling strengths. For pure Posner molecules we use, as a starting point, the phosphorus-phosphorus J-coupling strengths as calculated by Swift \textit{et al}. \cite{fisher2}. It should be noted, however, that these coupling constants depend on the fact that Swift \emph{et al}. assume the Posner molecules have symmetric configurations. Other Posner molecule configurations would alter and multiply the possible J-coupling constants and there is some evidence that Posner molecules prefer low symmetry states at room temperature \cite{shivang}. Agarwal \emph{et al}. recently published a detailed analysis of J-coupling constants in the context of Posner molecules of varying symmetries \cite{shivang2}. Both Swift \emph{et al}. and Agarwal \emph{et al}. estimate the coupling constants from theoretical first principles calculations \cite{fisher2,shivang2}. Phosphorus-phosphorus J-coupling strengths in adenine triphosphate, a source of the pyrophosphate used to assemble Posner molecules, have been reported as being approximately two orders of magnitude bigger \cite{jung} than the values calculated by Swift \emph{et al}. \cite{fisher2}. For this reason, and given the lack of experimentally verified parameters, we examine how phosphorus nuclear coherence and entanglement in Posner molecules depends on the size of the J-coupling coupling constants. \\
\\
Lithium coupling constants are also difficult to estimate accurately from the literature. For this reason the J-coupling constants between relevant atoms were calculated using ORCA \cite{neese}. Specifically, we performed DFT calculations using the pcseg-2 basis set \cite{jensen1} for all atoms, and the pcJ-2 basis set \cite{jensen2} - built specifically for the calculation of these coupling constants - for phosphorus and lithium atoms. The hybrid B3LYP exchange-correlation functional was used. Agarwal \emph{et al}. conclude in their paper that although J-coupling constants play an important role in Posner spin dynamics, it is ultimately the size of the spin system that has the greatest effect. We were interested to see whether our results confirmed their conclusion as we add the extra spins of the lithium isotopes \cite{shivang2}.  What is also interesting in the context of a discussion of the different effects of lithium isotopes on the spin dynamics of Posner molecules is that $^{6}\textrm{Li}$ and $^{7}\textrm{Li}$ have J-coupling strengths that vary to a measurable degree. Scalar or J-coupling is the indirect interaction of nuclear spins through their intermediate interaction with surrounding electrons. For lithium isotopes the different gyromagnetic ratios of the nuclei in question have an effect on the coupling strength, with larger gyromagnetic ratio of $^{7}\textrm{Li}$ resulting in larger J-coupling constants. Comparison of $^{6}\textrm{Li}$ and $^{7}\textrm{Li}$ gyromagnetic ratios gives $\gamma_7/\gamma_6=2.6$ \cite{spencer, gielen,gunther}. We have used this ratio to compare the effects of different lithium isotopes on nuclear coherence and entanglement in lithium substituted Posner molecules. 
\begin{figure*}[t!]
	\centering
	\includegraphics[scale=0.25]{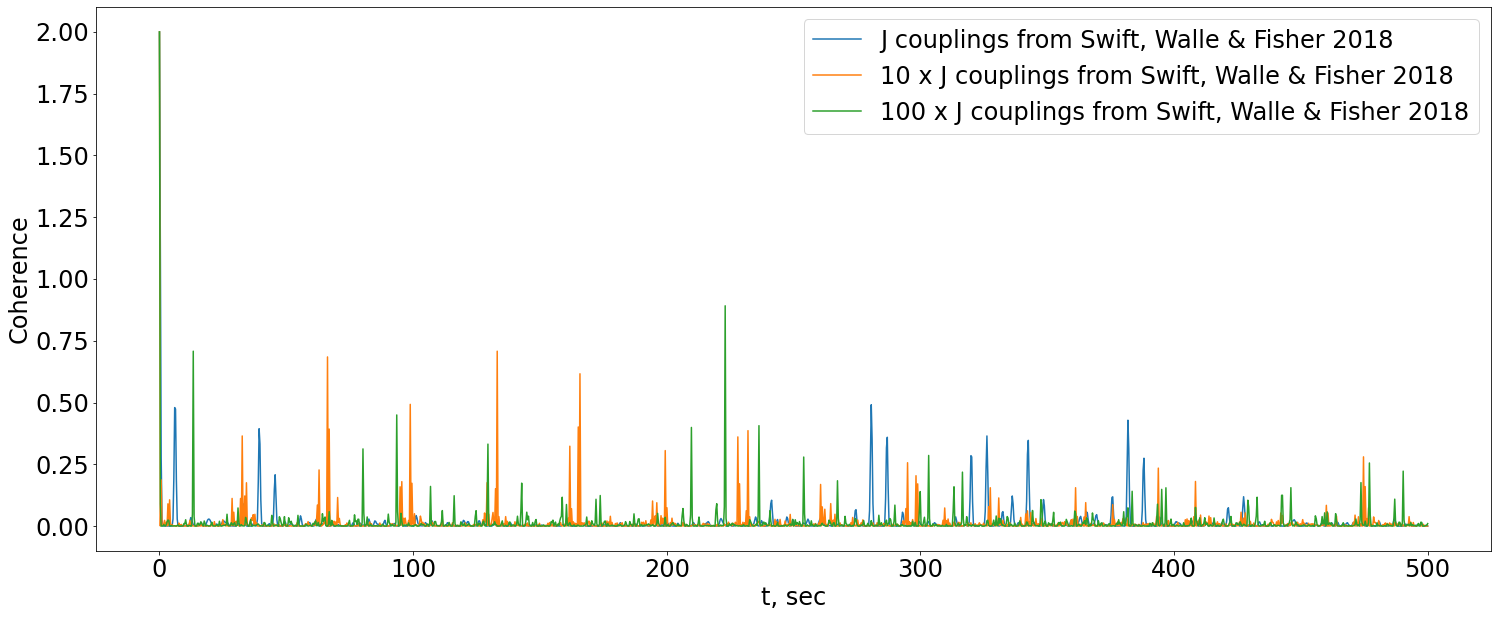}
	\includegraphics[scale=0.25]{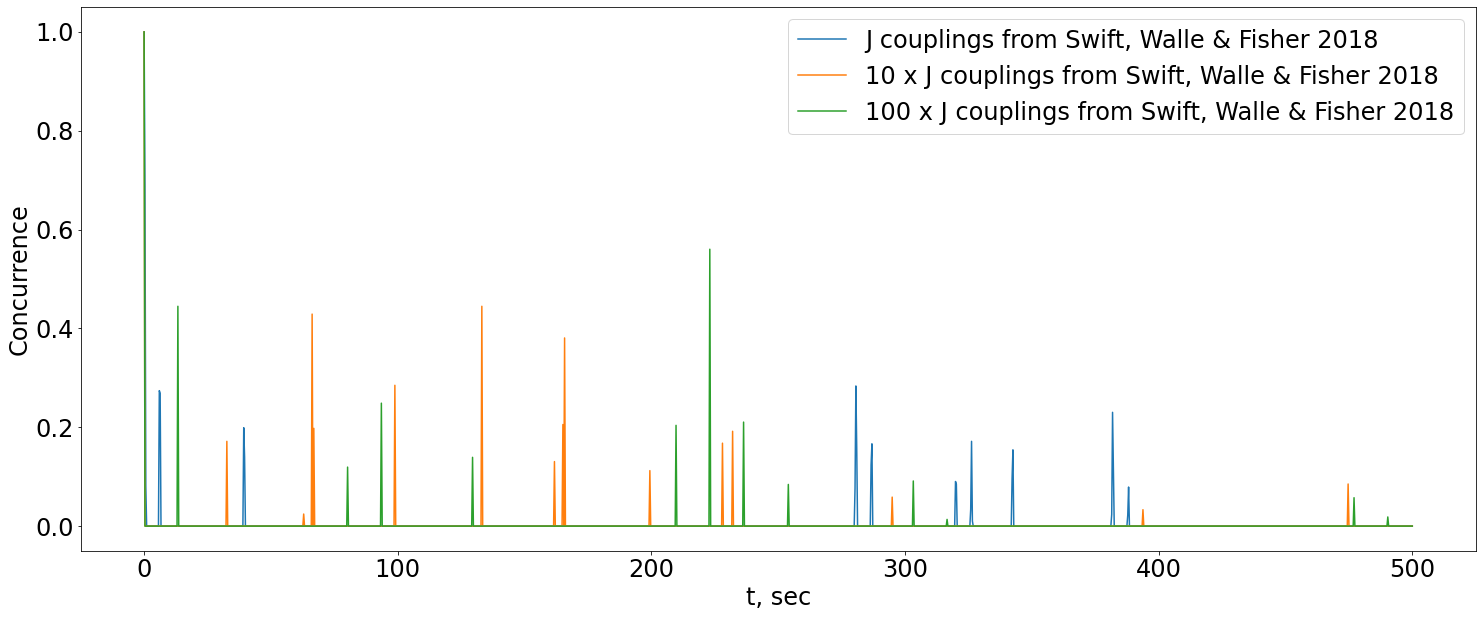}
	\caption{Dependence of coherence (top) and concurrence (bottom) on the J-coupling strengths from smaller to larger values: For coupling strengths of the order of those calculated by Swift \emph{et al}. \cite{fisher2} very little coherence and concurrence remain when the interactions of all phosphorus nuclei are considered. Both can be increased by increasing the J-coupling strength. Phosphorus-phosphorus J-coupling strengths in adenine triphosphate, a source of the pyrophosphate used to assemble Posner molecules, have been reported as being approximately two orders of magnitude bigger \cite{jung} than the values calculated by Swift \emph{et al}. \cite{fisher2} for Posner molecules with specific symmetries.}
\end{figure*}
\begin{figure*}[t!]
	\centering
	\includegraphics[scale=0.25]{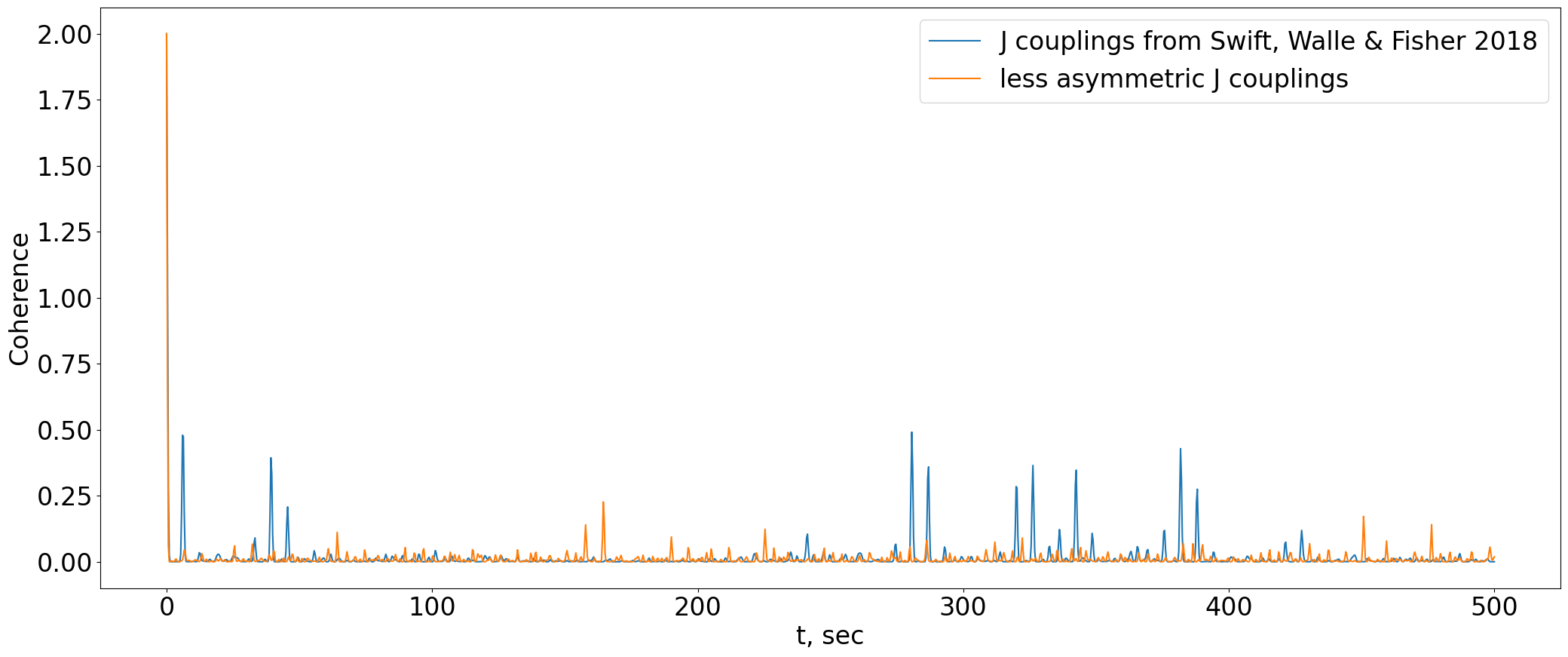}
	\includegraphics[scale=0.25]{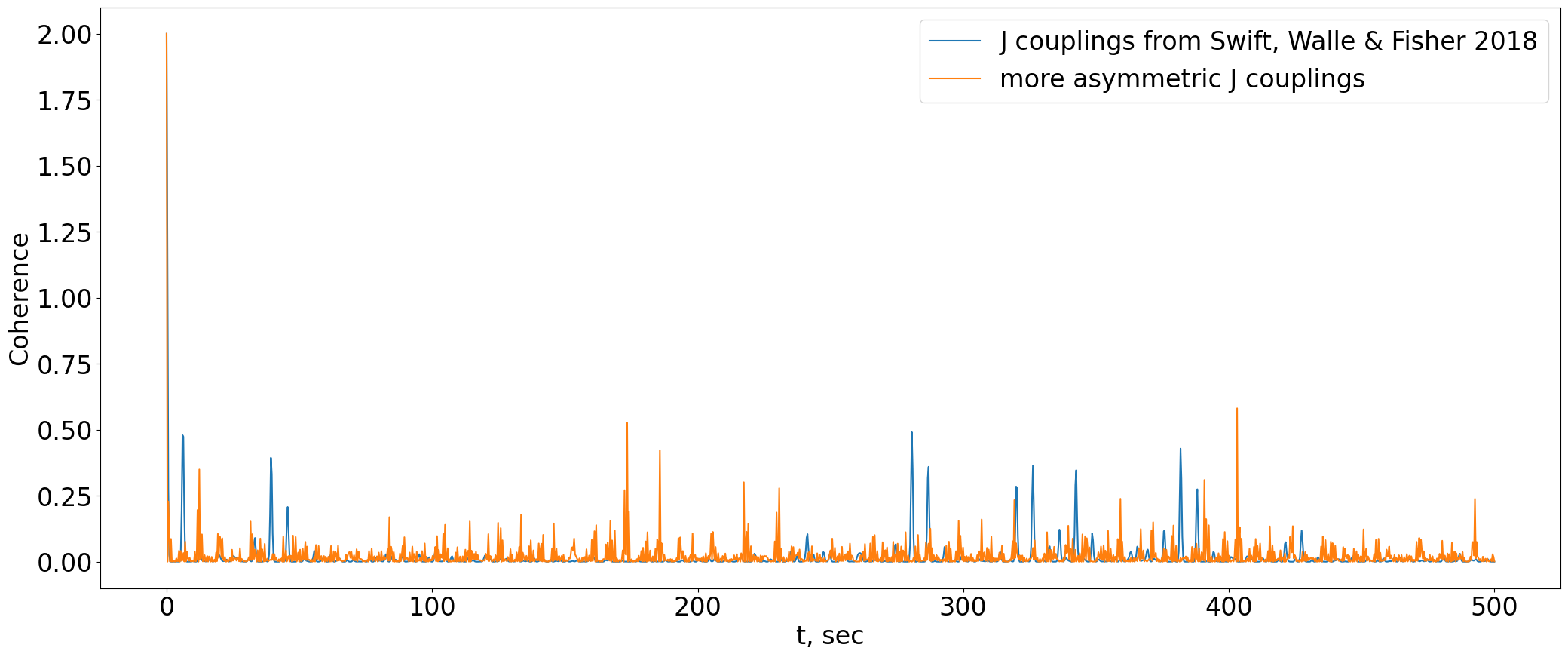}
	\caption{Dependence of coherence on the specific symmetries of the Posner molecules: We tried two different asymmetric configurations by varying the J-couplings. In the top graph, where the J-couplings are all different but both halves of the molecule are comparatively strongly coupled the coherence is attenuated compared to the symmetric case. In the bottom graph the asymmetry is more pronounced, with one side of the Posner molecule being very strongly coupled in comparison to the other half. Surprisingly, this caused an increase in coherence, possibly due to the strongly one-sided coupling effectively reducing the dimension of the Posner molecules.}
\end{figure*}
\begin{figure*}[t!]
	\centering
	\includegraphics[scale=0.25]{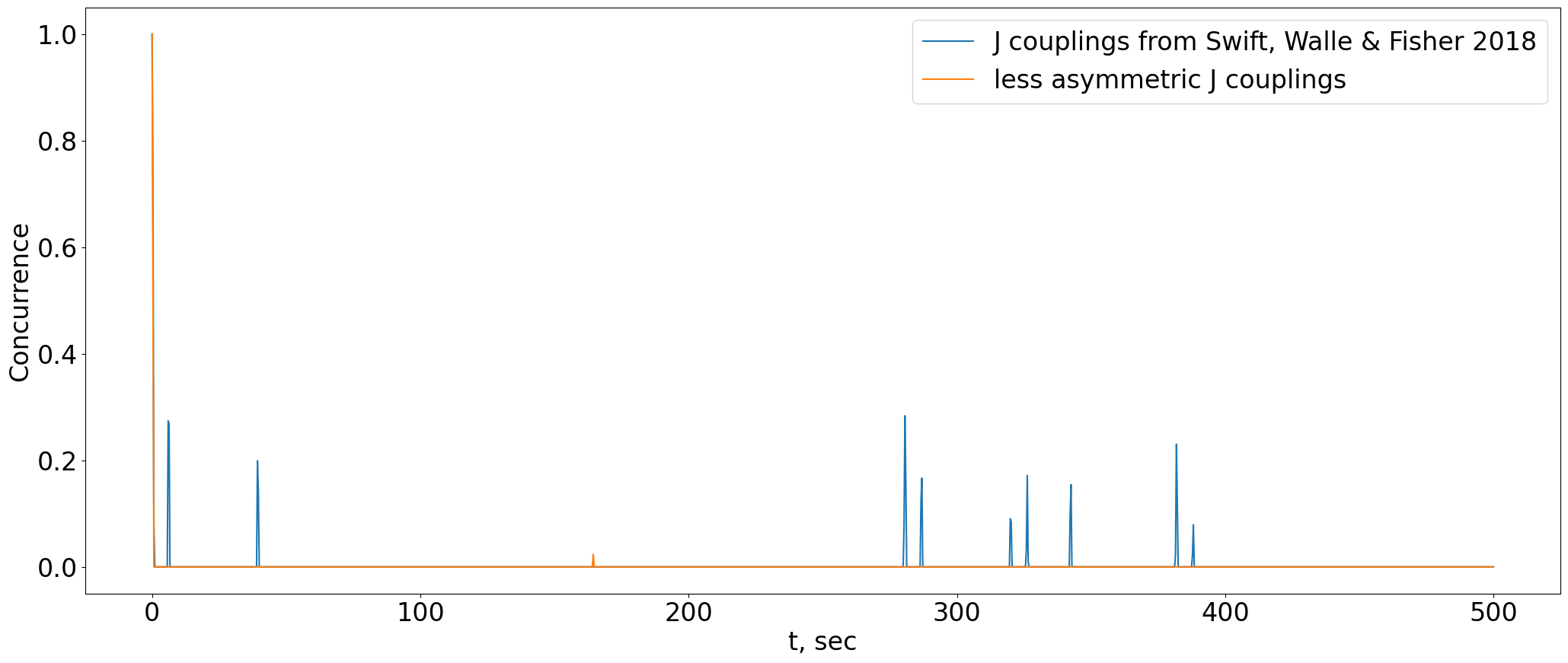}
	\includegraphics[scale=0.25]{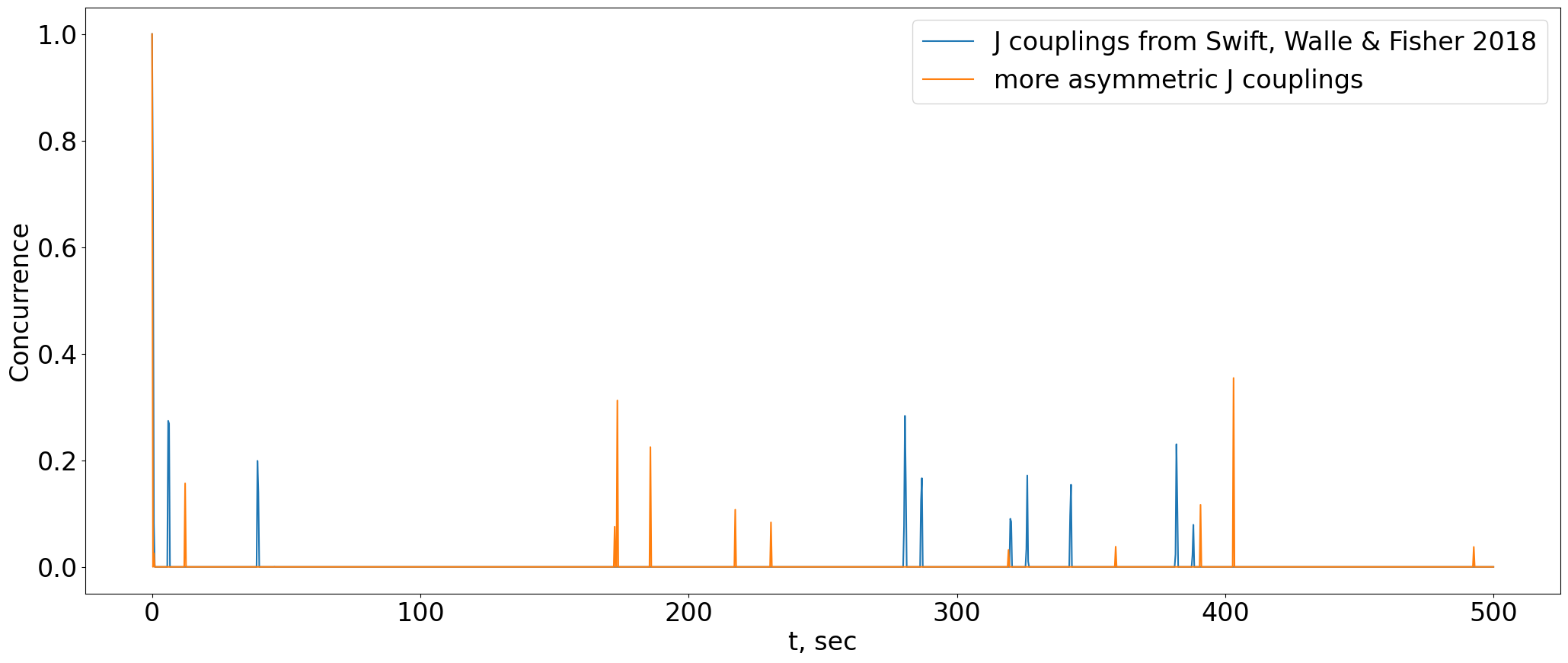}
	\caption{Dependence of concurrence on the specific symmetries of the Posner molecules: Similarly to the case for coherence, we tried two different asymmetric configurations by varying the J-couplings. In the top graph, concurrence is almost nonexistent. In this case the J-couplings are all different but both halves of the molecule are comparatively strongly coupled, with very weak concurrence between 100 and 200 seconds. These results appear to be partially in agreement with the conclusion in the recent paper by Agarwal \emph{et al} that on average symmetric molecules are expected to have a better
    entanglement yield \cite{shivang2}. However, when we arranged the J-couplings so that there is a very strongly coupled half of the Posner molecule (with respect to the entangled nuclei) then the concurrence is markedly increased. Very strong asymmetry in effect reduces the dimension of the Posner molecules, which appears to have the effect of increasing the entanglement.}
\end{figure*}
\begin{figure*}[t!]
	\centering
	\includegraphics[scale=0.25]{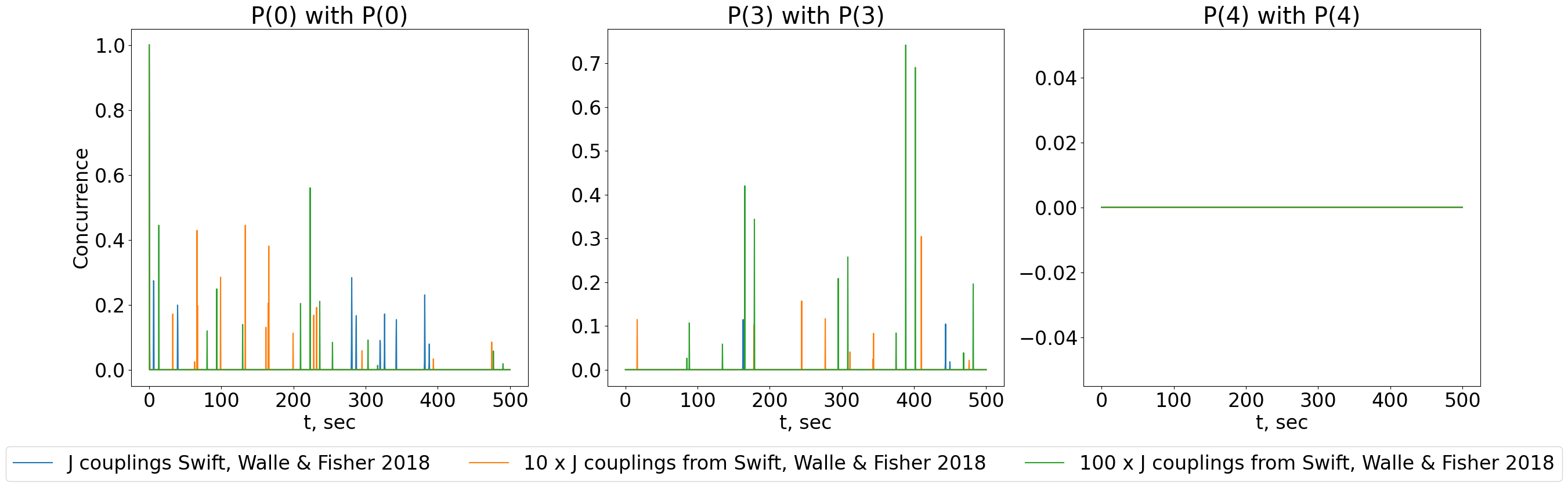}
	\caption{Entanglement transfer between the different phosphorus nuclei in a Posner molecule, where only one of the nuclei of each Posner molecule is initially entangled, P(0) with P(0): Entanglement is only transferred between the initial phosphorus P(0) and the furthest phosphorus P(3) in each of the entangled Posner molecules, regardless of J-coupling strength. There is zero entanglement seen between any of the other pairs, where P(4) and P(4) is given as an example.}
\end{figure*}
\begin{figure*}[t!]
	\centering
	\includegraphics[scale=0.25]{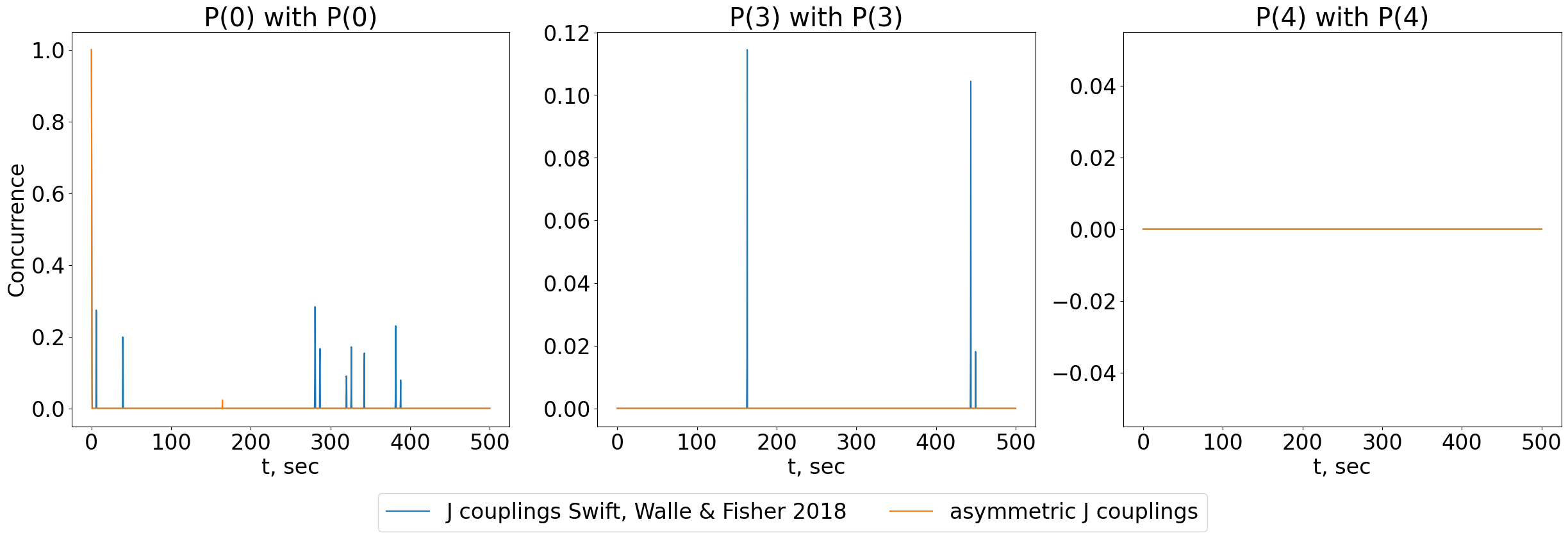}
	\caption{Dependence of entanglement transfer on symmetry constraints: Entanglement transfer depends on the symmetry of the molecule with the weakly asymmetric case (orange) having no entanglement transfer. There is zero entanglement seen between any of the other pairs, in both symmetric and asymmetric cases where P(4) and P(4) is given as an example, (the blue graph is beneath the orange).}
 \end{figure*}
\begin{figure*}[t!]
	\centering
	\includegraphics[scale=0.25]{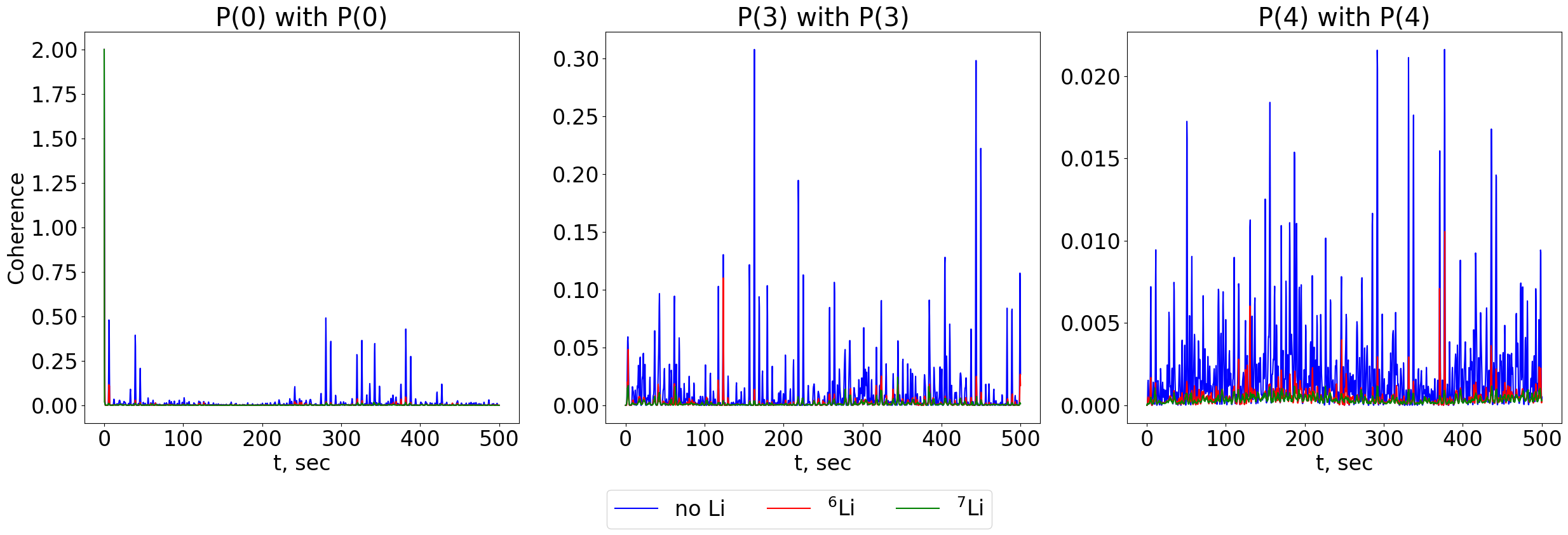}
	\includegraphics[scale=0.25]{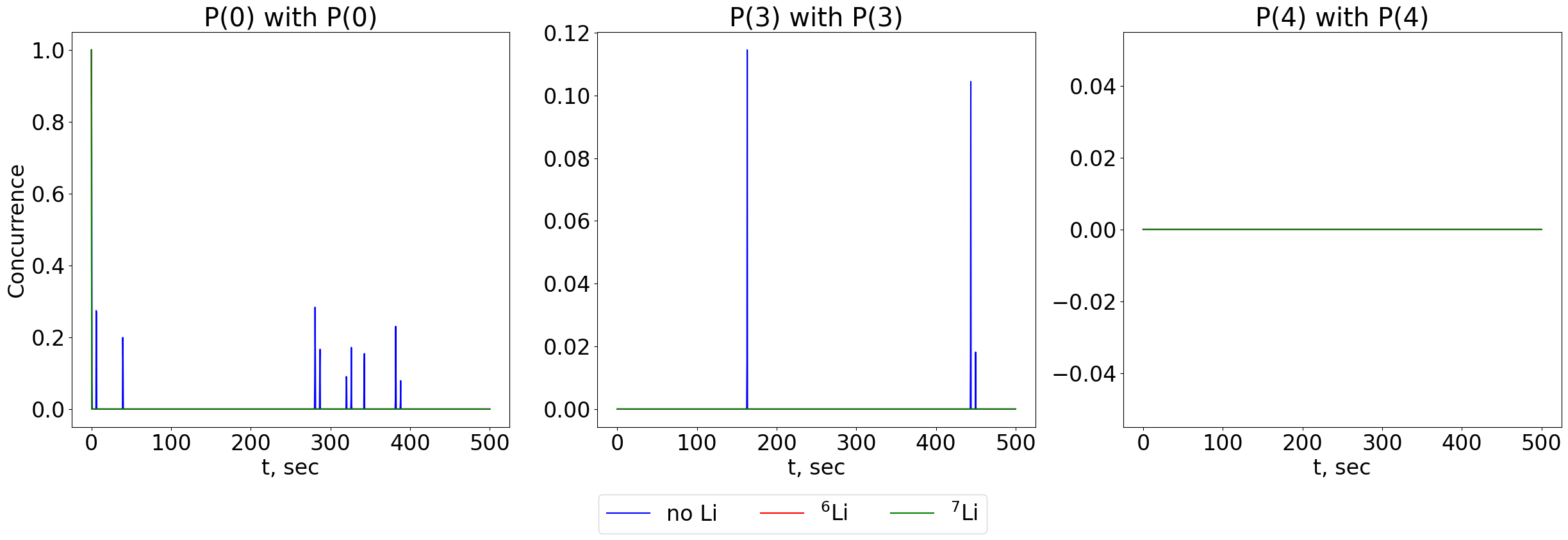}
	\caption{Effect of lithium isotopes on coherence (top) and entanglement (bottom) transfer: For the case of coherence we have increased the scale of the y-axis to illustrate how coherence is transferred differently between different nuclei. For two entangled spins initially P(0) and P(0) the maximum coherence is transferred to the furthest spins in each molecule, although there is a very small degree of coherence between other pairs, see for example P(4) with P(4) (with attention to the y-axis scale). Greatest coherence transfer is seen in pure Posner molecules (blue), followed by lithium 6 (red) and lithium 7 (green). However the overall coherence is still minimal. In the case of entanglement, only pure Posner molecules (blue) show any entanglement and entanglement transfer.}
\end{figure*}
\begin{figure*}[t!]
	\centering
	\includegraphics[scale=0.28]{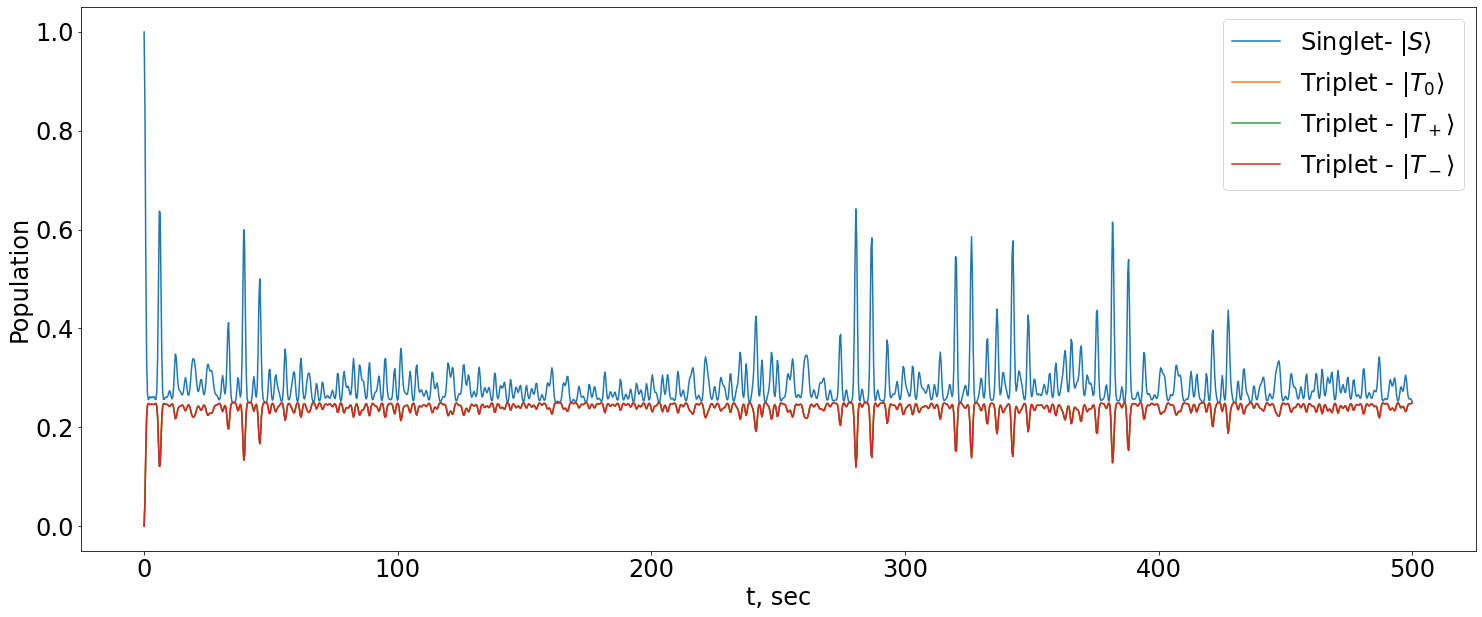}
	\includegraphics[scale=0.28]{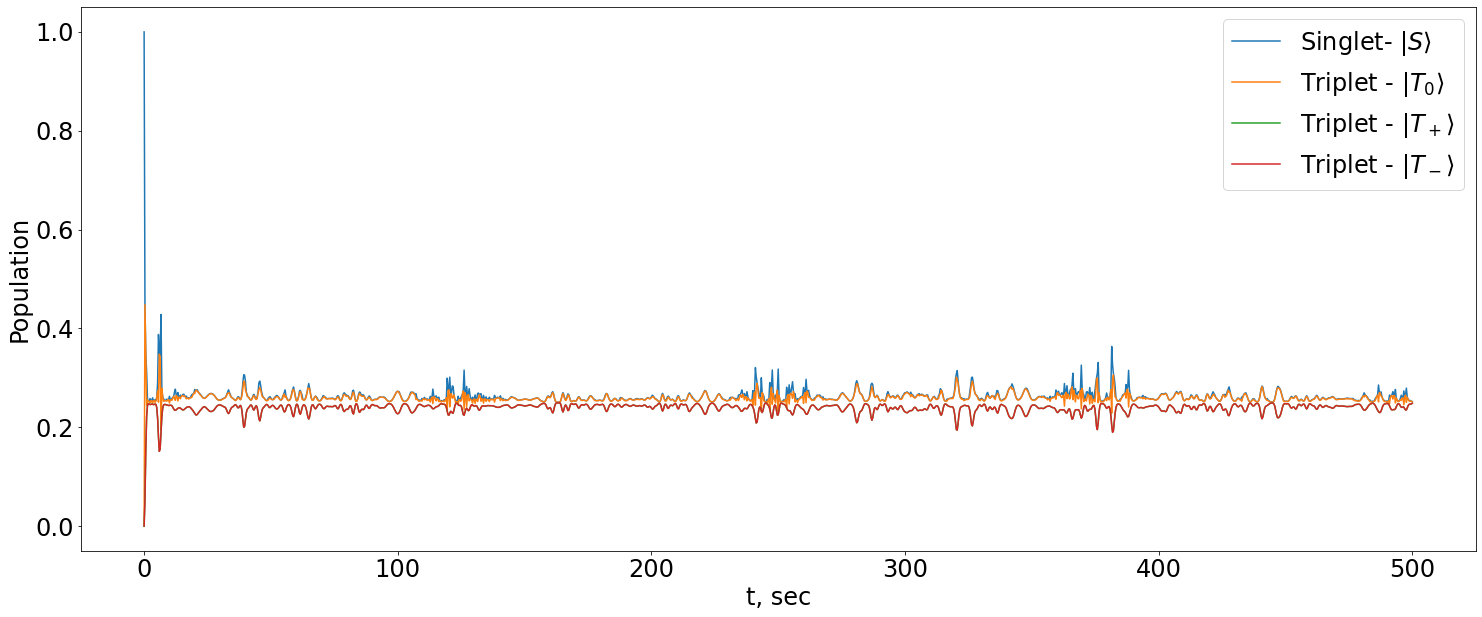}
	\includegraphics[scale=0.28]{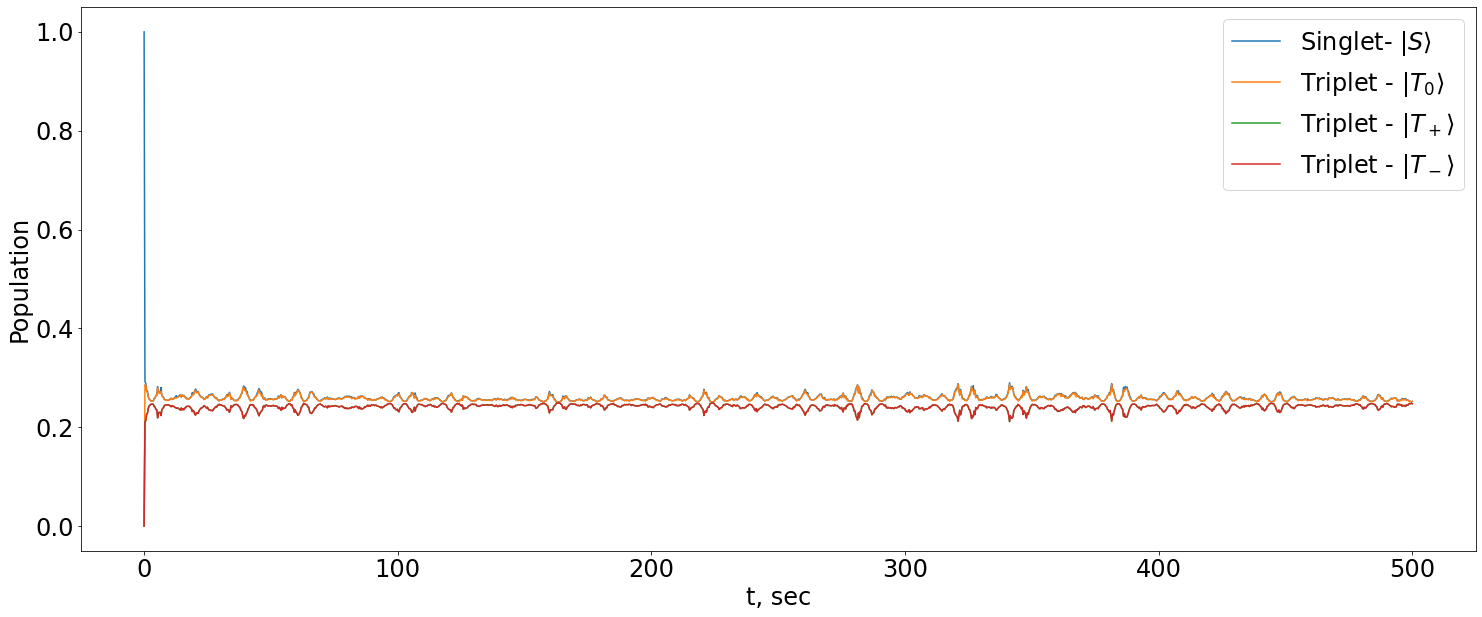}
	\caption{Effect of lithium isotopes on the spin dynamics: For pure Posner molecules (top graph) the three triplet states are degenerate and out of phase with the singlet state. Introducing lithium 6 (middle) and lithium 7 (bottom) causes the entangled triplet state to instead oscillate in phase with the entangled singlet state.}
 \end{figure*}

\section{Results}
\subsection{Pure Posner molecules}
Before considering phosphate nuclear entanglement in the context of lithium doped Posner molecules we consider the simplest case of an undoped Posner molecule to be clear as to what we mean by coherence or concurrence. Coherence, in this context, refers to the distance of a given state from a maximally mixed state. Concurrence corresponds to the probability of the entangled singlet state being above one half. In Figure 2 we show how increasing the strength of the J-couplings increases the coherence and concurrence. We have illustrated this by plotting the coherence and concurrence for J-coupling strengths beginning with those calculated by Swift \emph{et al}. and increasing in increments up to at least two orders of magnitude. There are now two papers in which Posner molecule J-coupling constants have been estimated using theoretical calculations \cite{fisher3,shivang2}. However, in the brain, ATP, the molecule from which phosphates are hypothetically assembled into Posner molecules, has been measured to have J-coupling strengths up to two orders of magnitude greater \cite{jung} than those calculated by Swift \emph{et al}. for Posner molecules. The J-coupling constants calculated by Swift \emph{et al}. \cite{fisher2} are also for a very specific Posner molecule configuration and it is unclear whether this is the energetically preferred configuration \cite{shivang}. In order to investigate what effect the symmetry of the molecule might play we plotted the coherence (Figure 3) and concurrence (Figure 4) for two asymmetric Posner configurations. We approximated the asymmetry by using different J-couplings. For the first configuration the two halves of the Posner molecules, with respect to the entangled nuclei, were comparably strongly coupled. For the second configuration there was one very strongly coupled half. In the weakly asymmetric case coherence and concurrence were both attenuated compared to the symmetric case. This would appear to be in agreement with the conclusion by Agarwal \emph{et al}. that on
average entanglement is better supported by symmetric molecules \cite{shivang2}. However, in the case of the very strongly asymmetric Posner molecule, the coherence and concurrence are surprisingly increased. This may be due to a combination of increased J-coupling strengths as well as an effectual reduction in dimension of the Posner molecule due to the unequal coupling strengths. The latter is potentially interesting given the suggestion that entanglement is increased in calcium phosphate dimers as compared to trimers \cite{shivang2}. Following the example of Player and Hore \cite{player} we also investigated how entanglement is transferred between the different phosphorus nuclei in a Posner molecule, where only one of the nuclei is initially entangled. Entanglement is only transferred between the initial phosphorus and the furthest phosphorus in each of the entangled Posner molecules, regardless of J-coupling strength, see Figure 5. We also investigated how entanglement transfer depends on the symmetry of the molecule with the weakly asymmetric case having no entanglement transfer whatsoever, see Figure 6.

\subsection{Doped Posner molecules}
It was noted that it is possible to replace calcium ions in calcium phosphate aggregates with other appropriate ions, such as sodium, magnesium or lithium \cite{fisher1,fisher2,mancardi}. Lithium is monovalent, therefore to replace a calcium ion in a pure Posner molecule requires two lithium ions \cite{fisher2}. As with pure Posner molecules the spin state oscillation depends on the J-coupling constants. In Figure 7 we consider how the different lithium isotopes change the coherence and concurrence of a pair of phosphorus nuclei. In the case of coherence, the two isotopes do have different effects, with more coherence in the $^{6}\textrm{Li}$ case. However the coherence in both cases is so small as to be almost negligible. In the case of concurrence, both $^{6}\textrm{Li}$ and $^{7}\textrm{Li}$ destroy any entanglement, as measured by singlet state population. There is also no entanglement transfer. These results are not promising with regards to Fisher's suggestion that lithium's mode of action is through being incorporated into Posner molecules. They also reiterate the conclusion in Agarwal \emph{et al}. that the size of the spin system is the main constraint on the entanglement \cite{shivang2}. \\
\\
What, if anything, does the incorporation of lithium do to the spin dynamics of entangled phosphorus nuclei. The answer might lie in a closer inspection of the Zeeman interaction. We investigated what effect the lithium isotopes have on the spin dynamics of the singlet and three triplet states, see Figure 8. For pure Posner molecules (top graph) the three triplet states are degenerate and out of phase with the singlet state. Introducing $^{6}\textrm{Li}$ and $^{7}\textrm{Li}$ causes the entangled triplet state to instead oscillate in phase with the entangled singlet state. This might be seen as analogous to the high field effect for a radical pair, when the hyperfine interaction is sufficiently smaller than the external magnetic field and the two non-entangled triplet states are separated in energy from the entangled states. Coherent mixing of the singlet and entangled triplet states in radical pairs can be driven by different Larmor precession frequencies, with the frequency of mixing related to the difference in gyromagnetic ratio \cite{woodward}. In the Posner analogy the Earth's magnetic field is at least three orders of magnitude larger than the J-coupling interaction. In addition to this, the small gyromagnetic ratio of $^{6}\textrm{Li}$ gives a precession frequency of $\approx$ 1970 Hz whereas $^{7}\textrm{Li}$ gives $\approx$ 5200 Hz, very close to phosphorus at $\approx$ 5420 Hz.\\

\section{Discussion}

\subsection{Relaxation pathways} 
In the previous sections we have considered only the coherent dynamics. For example we plotted the dynamics of lithium doped Posner molecules for up to 500 seconds (see Figure 8). If spin relaxation is taken into account then it must be acknowledged that both lithium isotopes have a quadrupolar moment that means they relax faster than the spin $\frac{1}{2}$ phosphorus nuclei. Lithium lifetimes vary in the literature; Fisher's original papers discussing entanglement in Posner molecules notes the difference in coherence lifetimes between solvated $^{6}\textrm{Li}$ ($\approx$ 5 minutes) and $^{7}\textrm{Li}$ ($\approx$ 10 seconds) while hypothesising that phosphorus lifetimes might be as long as 21 days \cite{fisher1, fisher3}. In their paper on the spin dynamics of Posner qubits, Player and Hore dispute this, invoking intramolecular dipole interactions to arrive at an estimate of 37 minutes \cite{player}. They also discuss a number of ways in which this lifetime might be significantly reduced, one of which is the replacement of calcium by other ions such as sodium \cite{player}. We add to this a possible relaxation pathway that is specific to the case of lithium isotopes, that is scalar relaxation. For a spin half nucleus A (phosphorus) coupled to a second nucleus B (lithium) that is undergoing fast quadrupolar relaxation, the fluctuating magnetic field associated with fast relaxing nucleus B offers an additional relaxation mechanism for nucleus A. This is known as scalar relaxation and is seldom taken into account due to the fact that it is most effective for nuclei that have similar Larmor frequencies \cite{chiavazza2013}. This dependence is given in the formula for the scalar relaxation lifetime
\begin{equation}\label{relaxscalar}
R_{\mathrm{sc}} = \frac{8\Pi^2 J^2}{3} I(I+1) \frac{\tau_{\mathrm{sc}}}{1+(\omega_B-\omega_A)^2 \tau_{\mathrm{sc}}^2},
\end{equation}
where $R_{\mathrm{sc}}$ is the relaxation rate of nucleus A, $J$ is the scalar coupling constant between A and B, $I$ is the spin of the quadrupolar nucleus B, $\tau_{sc}$ is the correlation time associated with the scalar relaxation and $\omega_A$ and $\omega_B$ are the respective Larmor frequencies, given by
\begin{equation}\label{larmor}
\omega_L = \frac{\gamma B_0}{2\pi},
\end{equation}
where $\gamma$ is the gyromagnetic ratio and $B_0$ is the magnetic field in question \cite{chiavazza2013}. Scalar relaxation is described by two types. We are interested in type 2, which occurs at low fields such as the geomagnetic field, which we here take to be $50\mu$T. For type 2 scalar relaxation $\tau_{sc}$ is the $T_1$ of the fast relaxing quadrupolar nucleus \cite{chiavazza2013}. What is of interest is that the Larmor frequencies of $^{6}\textrm{Li}$ and $^{7}\textrm{Li}$ differ to a large degree, whereas the Larmor frequencies of $^{31}\textrm{P}$ and $^{7}\textrm{Li}$ are close enough that scalar relaxation might offer a viable relaxation mechanism. To test this theory we calculated the relaxation rates and lifetimes for phosphorus nuclei in Posner molecules doped with either $^{6}\textrm{Li}$ or $^{7}\textrm{Li}$. We use relaxation time scales for lithium isotopes as given in Fisher's original paper on Posner molecules, although spin-relaxation times for lithium isotopes vary widely across the literature. Due to the similarities between the Larmor frequencies of $^{7}\textrm{Li}$ and $^{31}\textrm{P}$, scalar relaxation contributes an additional relaxation mechanism for the phosphorus nuclei in a Posner molecule, resulting in a phosphorus relaxation time of only seconds. Due to the large difference between the Larmor frequencies of $^{6}\textrm{Li}$ and $^{31}\textrm{P}$ the corresponding scalar relaxation lifetimes are at least five orders of magnitude greater.

\subsection{Electromagnetic noise}
It is often stated that a diagnostic test for the radical pair mechanism is the use of electromagnetic fields at frequencies equal to the singlet-triplet transition frequencies. In a previous paper we applied an open quantum systems approach to the radical pair mechanism to investigate transition operators and their related frequencies \cite{adams2018}. We reapply this model to the case of entangled phosphorus nuclei. In our model we only need consider the open systems description of a single Posner molecule as the entangled Posner molecules are coupled only through their initial conditions and are sufficiently separated to not interact further. The Hamiltonian of the open quantum system is the sum of a free term $H_0$ and an interaction term $H_{SB}$
\begin{equation}
H=H_0+H_{SB},
\end{equation}
where
\[
H_0=H_S+H_B.
\]
We will describe the dynamics in the interaction picture, in which both state vectors and operators evolve in time. The system, in this case for a pure Posner molecule, includes the six phosphorus nuclear spins of the Posner molecule and how these interact with the Earth's magnetic field (Zeeman effect) and each other (J-coupling). This is given by Equation (3). This system then interacts with a bath given by
\begin{equation}\label{bathham}
H_B=\sum_{n}\omega_{n}a^\dagger_{n}a_{n},
\end{equation}
where the $\omega_n$ are the frequencies of the $n\mathrm{-th}$ bosonic operator, $a^\dagger_n$ is the creation operator and $a_n$ is the annihilation operator. Each Posner molecule of the entangled pair interacts with a separate but identical bath. The interaction Hamiltonian for one Posner molecule can thus be written as
\begin{equation}
H_{SB}=\sum_{k}\sum_{n} (g_{n,k}a_{n}+g^*_{n,k}a_{n}^{\dagger})\otimes(\alpha_k S^k_x+S^k_z),
\end{equation}
where the index $k$ keeps track of the different phosphorus nuclei and $S^k_x$ or $S^k_z$ represent dissipation and decoherence respectively, with $\alpha_k\geq0$ a model parameter weighting the extent to which $S^k_x$ and $S^k_z$ contribute. What is of interest in a discussion of the entanglement are the transitions between possible states. The transition operators result from the decomposition of the operator $V^k=\alpha_k S^k_x+S^k_z$ in the basis of the eigenoperators of the diagonalised system Hamiltonian $H_S$. They are found using
\[
[H_S,V^k_q]=-\omega_{k,q} V^k_q \quad\mathrm{and}\quad [H_S,V_q^{k,\dagger}]=\omega_{k,q} V_q^{k,\dagger},
\]
where $q$ here labels the number of transition operators and the transition frequencies corresponding to each operator $V^k$, $\omega_{k,q}\geq 0$, are expressed in terms of the parameters of the system Hamiltonian, that is the magnetic field and the coupling constants \cite{adams2018}. A transition frequency equal to zero corresponds to decoherence in the system, which is found using
\[
[H_S,V^k_0]=0.
\]
Given that the transition frequencies depend on the specific parameters of the system Hamiltonian, they will vary according to the relative strengths of the Zeeman and scalar coupling terms. For the radical pair mechanism, which is concerned with electron spin, hyperfine coupling strengths range from kHz to MHz \cite{rodgers}. This gives transition frequencies that range from kHz to MHz, which seems consistent with the fact that the avian compass is disrupted by broadband electromagnetic noise \cite{mouritsen,adams2018}. For the Posner molecule nuclear spin states the transition frequencies vary according to the strength of the Zeeman and J-coupling terms. In particular the two entangled states, which are unaffected by the external magnetic field, have transition frequencies which reflect the J-coupling constants, which are of the order of Hz. This is potentially interesting in light of the fact that the brain emits electromagnetic radiation of the order of Hz, colloquially known as `brain waves' or neural oscillations. It is unclear whether this radiation is directly associated with entangled calcium ion production. However, given that the interconversion of entangled states is of similar frequencies, the background electromagnetic radiation generated by the brain (and other organs) should be taken into account as a possible source of noise or driven interconversion when discussing the spin dynamics of Posner molecules.
\begin{figure*}[t!]
	\centering
        \includegraphics[scale=0.3]{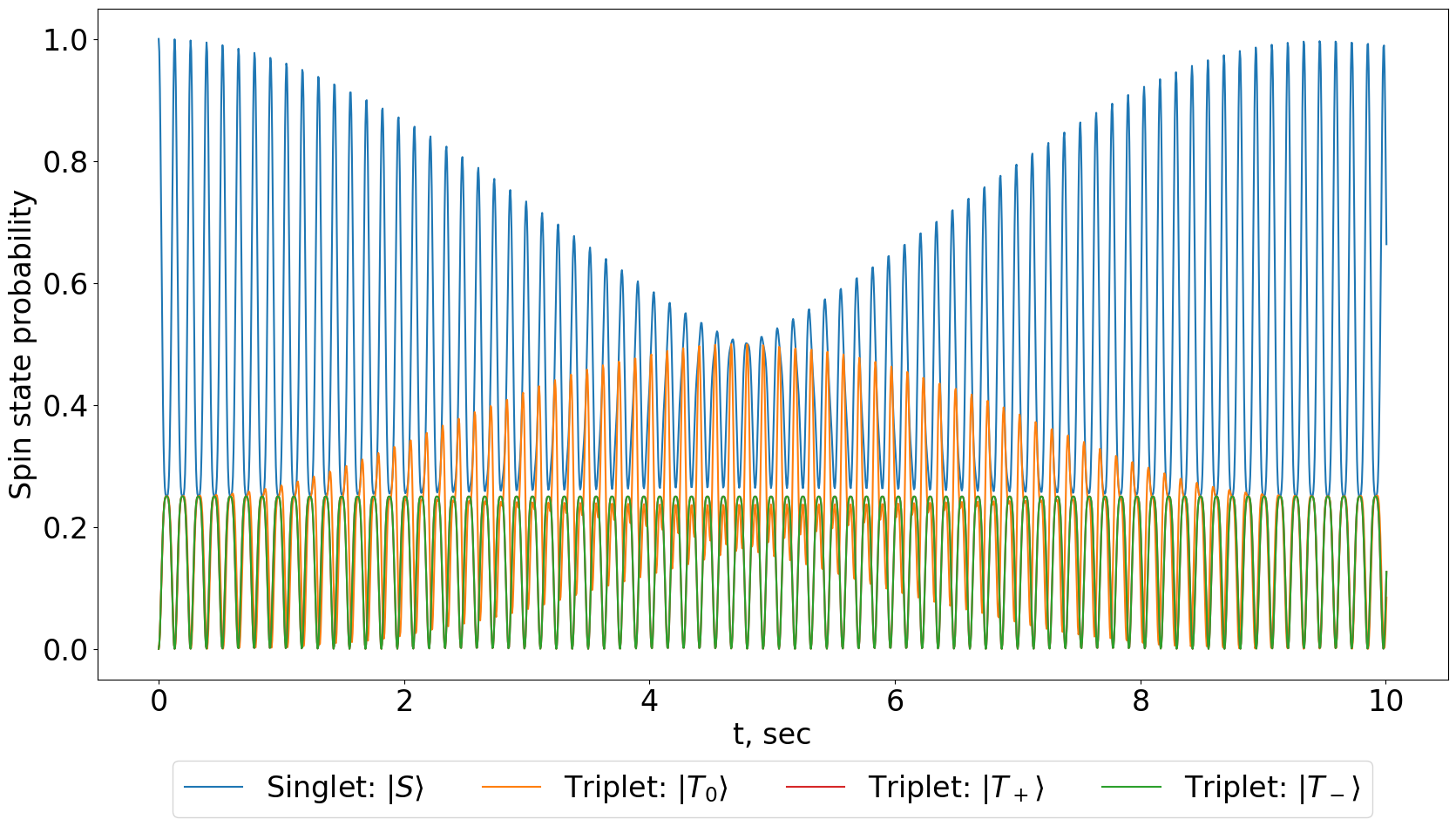}
	\includegraphics[scale=0.3]{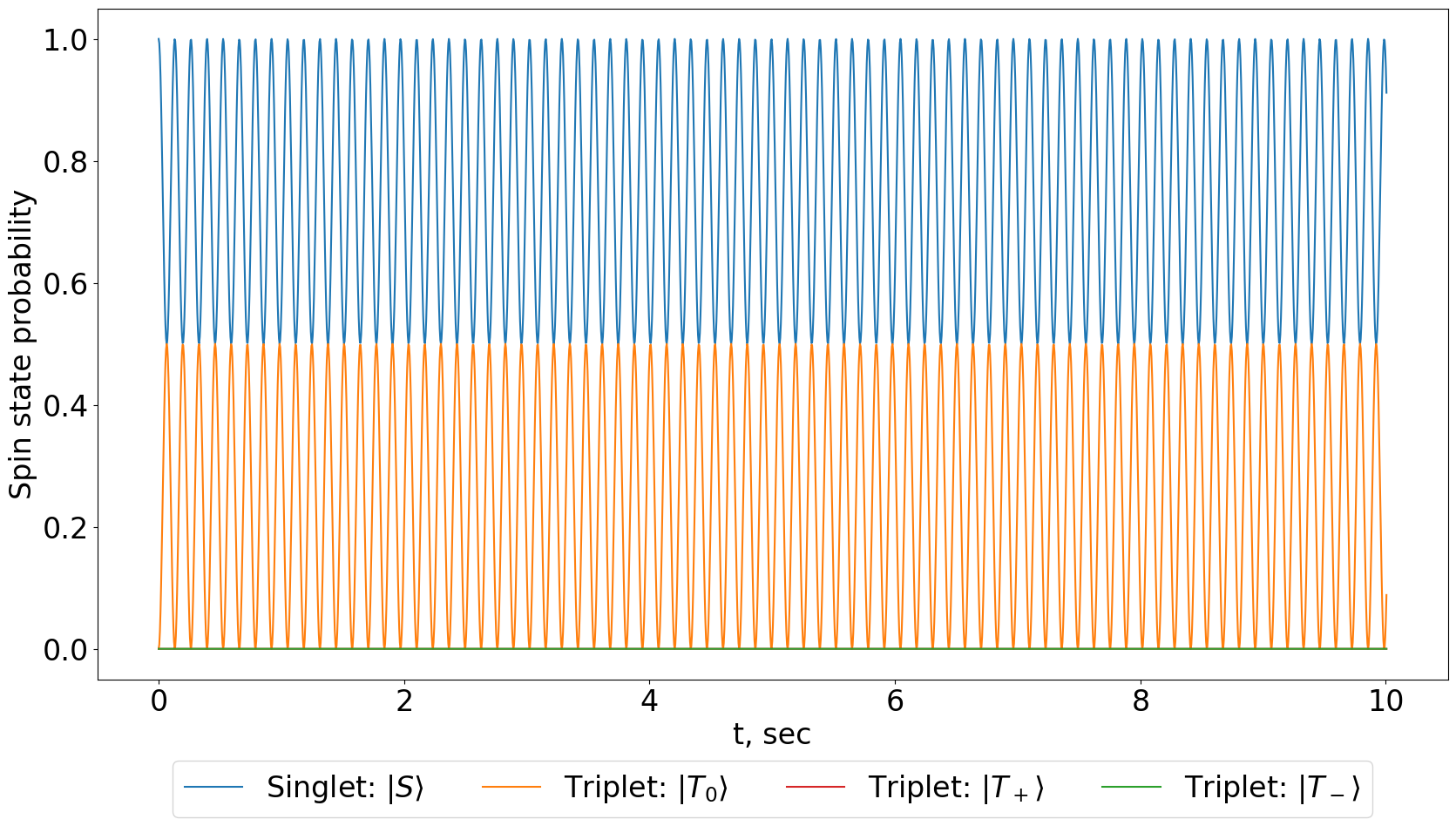}
	\caption{Different spin states of the entangled phosphates bound with hydrogen: When the external magnetic field is of comparable strength to the J-coupling there is mixing between all four of the spin states, (top). Note the red and green states are degenerate. However, for an external magnetic field that is much stronger than the J-coupling, as is the case for Posner molecules in an Earth strength field, then the two non-entangled triplet states are sufficiently separated in energy from the entangled states to prevent mixing. The result is an entangled subspace, where only the entangled states (blue and orange) oscillate and the non-entangled triplets (green and red are degenerate) show a straight line (bottom).}
\end{figure*}
\begin{figure*}[t!]
	\centering
	\includegraphics[scale=0.3]{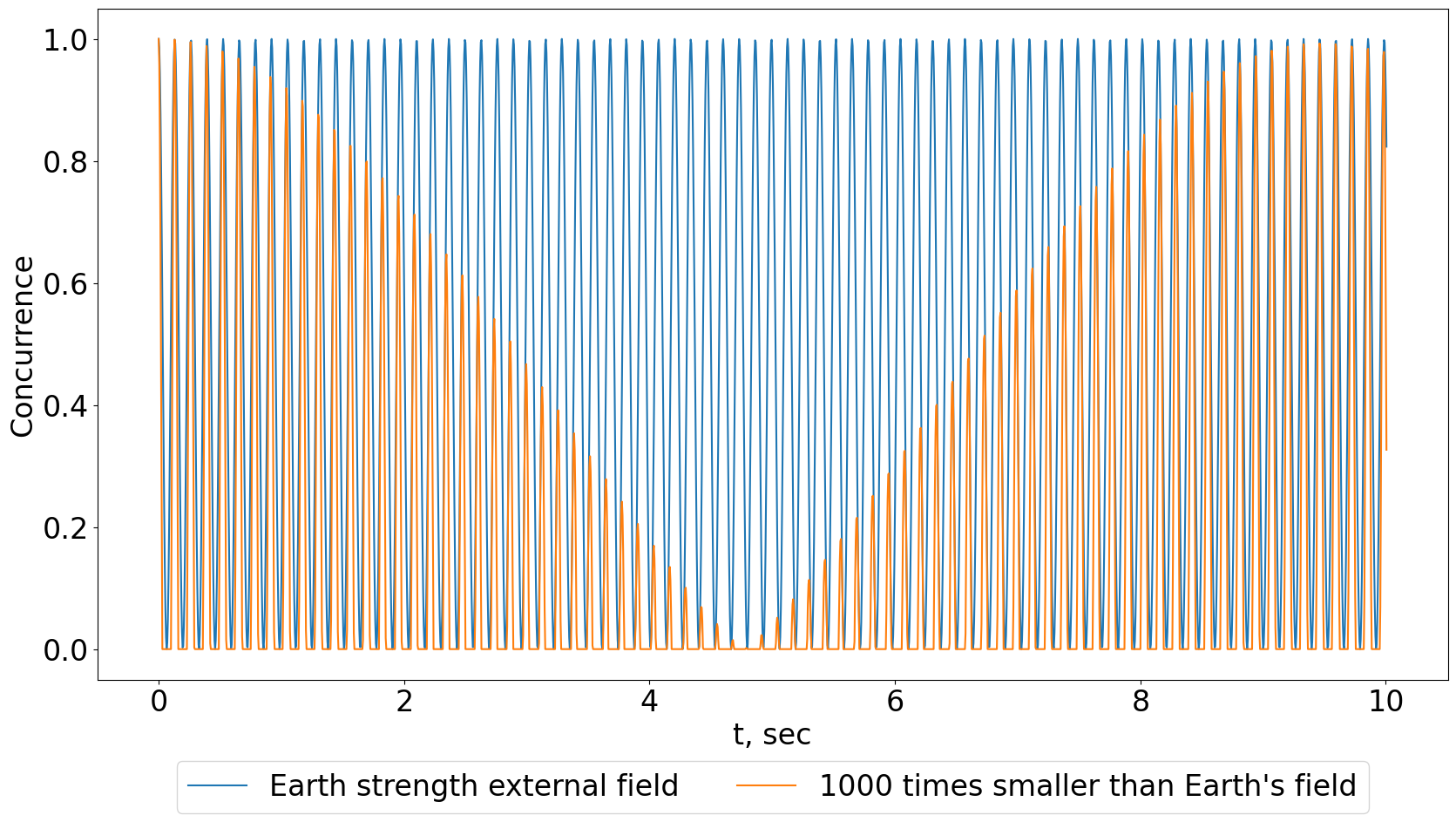}
	\caption{Comparison of entanglement for different strengths of the external field with respect to the J-couplings. The blue graph shows how entanglement is increased for an external field, such as the Earth's field, that is three orders of magnitude greater than the J-coupling strengths. The orange graph shows the decreased entanglement for an external field that is of the order of the J-couplings. Our measure of entanglement is concurrence which looks at singlet character. However there is also an entangled triplet state. Including this in the measure of entanglement should give us a fully entangled subspace.}
\end{figure*}
\subsection{Entangled subspaces}
In Fisher's original Posner molecule hypothesis, the entanglement is important in the context of quantum to biological transduction: how do the quantum effects result in measurable biological outcomes. Fisher contends that the Posner molecule entanglement results in modified molecule binding, melting and free calcium ion release \cite{fisher1}. If entanglement is indeed important with regards to Posner molecules and neural activation, how might biological systems have maximised this quantum resource. Agarwal \emph{et al}., for instance, suggest that different forms of calcium phosphate, for example dimers rather than trimers, are better suited to neural processing, having very long lived entanglement \cite{shivang2}. We suggest here that the parameters supplied by the particular environmental context -- the specific values of the Zeeman effect in the Earth's field relative to the strength of the J-coupling constants --  act to naturally enhance the entanglement. In a paper investigating entanglement in radical pairs, Tiersch \textit{et al}. suggest that one way in which to enhance the entanglement lifetime is the application of appropriate magnetic fields. This would create a maximally entangled subspace by increasing the energy separation of the two non-entangled triplet states $T_+ =\ket{\uparrow\uparrow}$ and $T_- =\ket{\downarrow\downarrow}$ \cite{tiersch}. We were interested in how this could apply in the case of nuclear spin entanglement. Entangled phosphates are hypothesised to be created by the hydrolysis of pyrophosphate \cite{fisher1}. Before these entangled phosphates bind to spin zero calcium and form Posner molecules, they might also end up binding to hydrogen \cite{fisher1}. Hydrogen has a large gyromagnetic ratio. We were interested to see whether hydrogen binding to phosphates could instead be beneficial to enhancing the entanglement, by creating an entangled subspace. In Figure 9 we demonstrate what happens to the different spin states of the entangled phosphates bound with hydrogen. When the external magnetic field is of comparable strength to the J-coupling then there is mixing between all four of the spin states. However, for an external magnetic field that is much stronger than the J-coupling, as is the case for Posner molecules in an Earth strength field, then the two non-entangled triplet states are sufficiently separated in energy to prevent mixing. The result is an entangled subspace, where only the entangled states mix. In Figure 10 we illustrate this using concurrence as a measure of entanglement. The usefulness of this increased entanglement is debatable. Indeed, a paper by Eisert \emph{et al}. argues that in a diffusion model the loss of position information can degrade entanglement considerably \cite{plenio}. However, in the radical pair mechanism, for example, it is less the entanglement than it is the spin state that is important. Singlet and triplet states have differential reactivity. In our example of nuclear spin entanglement, increased entanglement also means increased singlet state. If this increased singlet yield plays a role in any biologically relevant chemical reaction, then the high field effect we describe here may be functionally important. Phosphorus is also found within cell membranes, which are composed of phospholipids. In this case the binding of hydrogen to fixed phosphates may possibly exploit this entanglement `distillation'. Entangled subspaces in the context of nuclear spin dynamics could also give insight into ways of enhancing entanglement in spin models of quantum computers, where the spins are fixed rather than diffusive. 

\section{Conclusion}
Our conclusions are tentative given the lack of definitive parameters in the context of Posner molecules, both with or without lithium. Indeed, what this highlights is the importance of parameters in quantum biology. For instance we demonstrated that across a viable range of J-coupling constants, increased coupling strength increased phosphorus nuclear coherence and concurrence in Posner molecules. Given that these coupling strengths depend on the symmetry of Posner molecules and that this symmetry is disputed, more research remains to be done to determine the relevant parameters. For lithium-doped Posner molecules our results do not support the hypothesis that different lithium ions differently influence phosphorus nuclear coherence or concurrence. Although there is a marginal difference between the isotopes, both isotopes result in almost negligible coherence and concurrence. This conclusion appears to be in agreement with the conclusion by Agarwal \emph{et al}. that is is primarily the number of spins in the system that attenuate the entanglement \cite{shivang2}. What is potentially interesting is that lithium does change the spin dynamics of the correlated phosphorus nuclei. If this were to translate to spin-dependent binding and thus free calcium ion production, then an argument could be made for lithium modulating the production of free calcium and thus neural excitability. A way to test this might be to confirm that treatment with lithium changes levels of free calcium and phosphate ions, and that this holds to different degrees for different isotopes. There is some evidence to suggest that administration of lithium does indeed have an effect on serum concentrations of these ions, though the mechanism remains debatable \cite{dorflinger,woo}. \\
\\
While our results for coherence and concurrence in lithium substituted Posner molecules remain inconclusive, the spin dynamics highlight a potentially interesting phenomenon. When entangled phosphates bind to hydrogen instead of calcium, it is proposed that quantum effects are destroyed. However, the fact that the external magnetic field is much larger than the scalar spin coupling allows for an effect analogous to the high-field effect in the radical pair, where mixing occurs only between the entangled states. In effect the Earth's magnetic field supplies an entangled subspace. This means that any travel outside of this magnetic field, such as space exploration and settlement will have to factor in these changes to the Zeeman splitting and the physiological implications thereof. This has further implications outside of biology, for example in the use of quantum computers that might use spin entanglement as a resource. Phosphates and phosphorylation are also ubiquitous in biological systems, and entangled phosphates may play a role outside of their incorporation into Posner molecules. Smolin, for instance, hypothesises how Fisher's theory of entangled phosphates might be combined with ideas from spin quantum computing and applied to the biological context of cell membranes, which are composed of phospholipids \cite{smolin}. While the viability of this remains to be seen, the parallel between phosphorus nuclear spin in the quantum computing and biological case -- especially given the potential for entanglement preservation -- leads us to the conclusion that the topic deserves further attention.

\end{document}